\documentclass{svproc}
\usepackage{amssymb}
\usepackage{amsmath}
\usepackage[T1]{fontenc}
\usepackage{graphicx}
\usepackage{url}

\begin{document}
\mainmatter              
\title{High Energy States of Recurrent Chaotic Trajectories in Time-Dependent Potential Well}

\author{Matheus S. Palmero\inst{1,2} \and Flavio H. Graciano\inst{3} \and
Edson D. Leonel\inst{4} \and Juliano A. de Oliveira\inst{5}}
\authorrunning{Palmero et al.} 

\institute{
Instituto de Ciências Matemáticas e de Computação, Universidade de São Paulo, Av. Trabalhador São Carlense 400, São Carlos, 13566-590, Brazil\\
\email{palmero@usp.br}
\and
Potsdam-Institut für Klimafolgenforschung, Telegrafenberg A56, Potsdam, 14473, Germany
\and
Departamento de Física, Universidade Estadual Paulista, Av. 24 A, Rio Claro, 13506-900, Brazil
\and
Instituto Federal do Sul de Minas Gerais, Av. Maria da Conceição Santos, 900, Pouso Alegre, 37550-000, Brazil
\and
Faculdade de Engenharia, Universidade Estadual Paulista, Av. Profa. Isette Corrêa Fontão, 505, São João da Boa Vista, 13876-750, Brazil
}

\maketitle              

\begin{abstract}
In this numerical study, recurrence quantification analysis of chaotic trajectories is explored to detect atypical dynamical behaviour in non-linear Hamiltonian systems. An ensemble of initial conditions is evolved up to a maximum iteration time, and the recurrence rate of each orbit is computed, allowing a subset of trajectories exhibiting significantly higher recurrences than the ensemble average to be identified. These special trajectories are determined through a suitable statistical distribution, within which peak detection reveals the respective initial condition that is evolved into a highly recurrent chaotic orbit, a phenomenon known as stickiness. By applying this methodology to a model of a classical particle in a time-dependent potential well, it is demonstrated that, for specific parameter values and initial conditions, such recurrent chaotic trajectories can give rise to transient high-energy states.

\end{abstract}

\section{Introduction}
\label{sec:intro}
Potential wells represent a fundamental model in classical mechanics, widely employed to study dynamical phenomena ranging from atomic-scale physics to macroscopic oscillatory systems \cite{Landau1976,Pollak1981A,Chiacchiera2016,Udani2018}. Classical examples such as particles confined within static or time-dependent potential barriers and wells have provided deep insights into chaotic dynamics, diffusive processes, and energy transfer mechanisms. Among these classical models, the Fermi–Ulam accelerator \cite{Fermi1949,Ulam1961,Leonel2004} and the bouncing-ball model (also known as the bouncer model) \cite{Lichtenberg1980,Leonel2008} stand out, illustrating rich dynamical behaviours including chaos and periodicity in response to oscillating boundaries or potentials.

Introducing periodic oscillations into Hamiltonian systems can drastically modify their dynamical properties, resulting in mixed-type phase spaces that contain both chaotic regions and stability islands \cite{Lichtenberg1992}. These mixed phase spaces characteristically exhibit intricate dynamical structures, such as invariant curves, periodic islands, and chaotic seas, shaped by system parameters and initial conditions. The interplay between stable and unstable dynamics, exemplified by the coexistence of periodic and chaotic orbits, forms the basis for complex transport phenomena, such as the intermittent trapping known as stickiness \cite{Zaslavsky2002,Wiggins2013}.

Recurrence analysis provides a robust tool to investigate the intricate temporal dynamics of chaotic trajectories. Techniques like Recurrence Plots (RPs) \cite{Eckmann1987} and Recurrence Quantification Analysis (RQA) \cite{Marwan2007,MarwanWebber2015} systematically quantify how trajectories revisit previously visited regions of phase space over time. The recurrence rate, in particular, measures the frequency of these returns, enabling identification and characterisation of distinct dynamical behaviours, including periodic motions, quasi-periodic states, chaotic exploration, and especially sticky dynamics. Recent advances demonstrate that RR effectively distinguishes sticky orbits from purely chaotic orbits in mixed phase spaces \cite{Palmero2022}.

In this paper, we perform a comprehensive recurrence analysis of chaotic trajectories confined in a time-dependent potential well. Employing an ensemble approach, we systematically identify specific initial conditions (ICs) that yield highly recurrent chaotic trajectories. These trajectories experience prolonged states of high energy, despite originating from low-energy initial conditions. Through detailed numerical experiments, we characterise the stickiness behaviour observed near invariant structures in phase space and elucidate the conditions under which these trajectories exhibit such remarkable energy excursions.

The paper is structured as follows: Section\ \ref{sec:model} introduces the potential well model. Section\ \ref{sec:methodology} describes the Ensemble Recurrence Analysis (ERA) method. Section\ \ref{sec:results} presents results, focusing on highly recurrent trajectories and stickiness behaviour. Finally, Section\ \ref{sec:conclusions} summarises findings and outlines future research perspectives.

\section{Model}
\label{sec:model}

\begin{figure}
    \centering
    \includegraphics[scale=0.21]{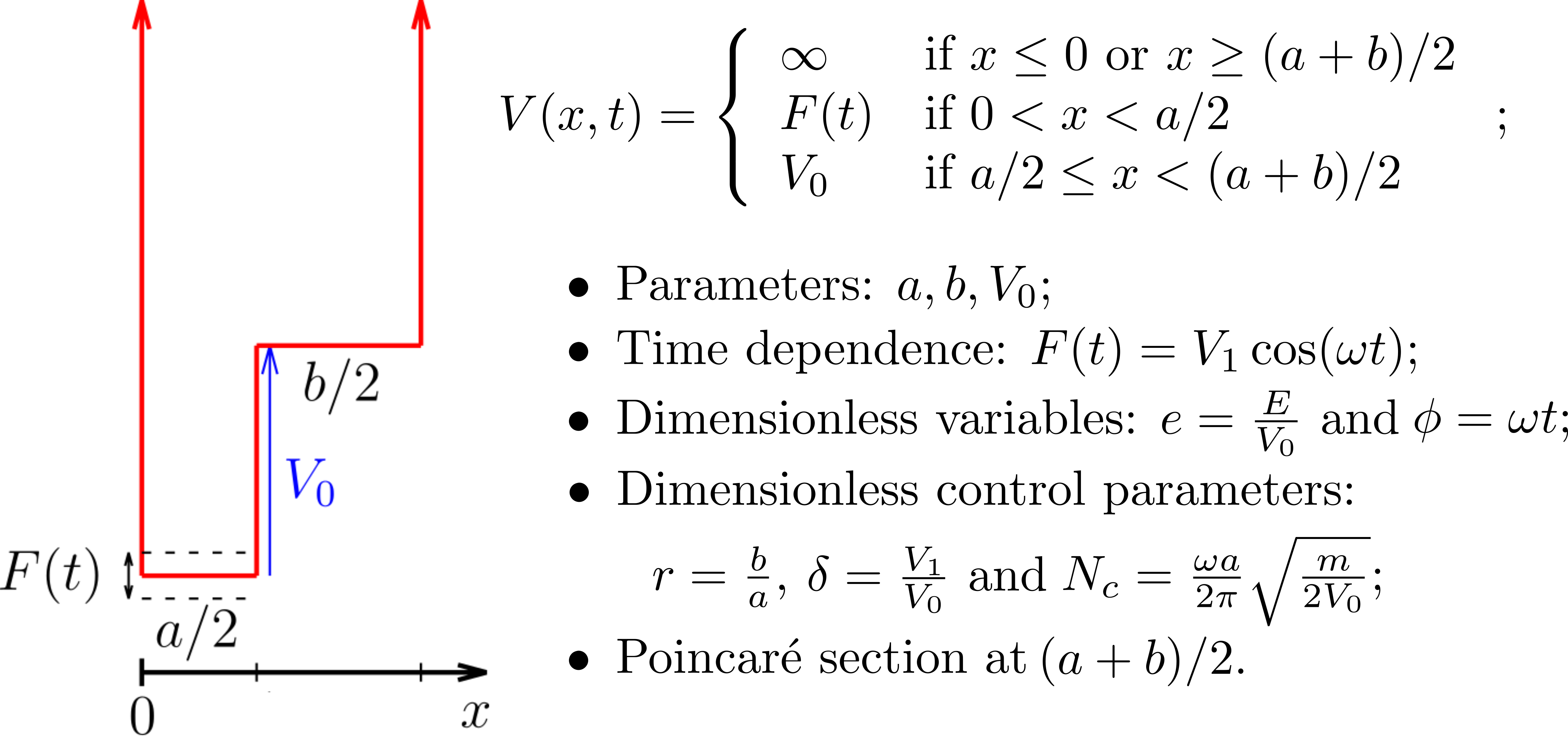}
    \caption{Schematic representation of the time-dependent potential well model, along with details on the potential energy, parameters, variables and defined Poincaré section.}
    \label{fig:model}
\end{figure}

We consider a classical particle confined within a one-dimensional, time-dependent potential well whose bottom oscillates periodically in time. The system consists of two regions: one with a fixed potential barrier and another with an oscillatory boundary, as illustrated in Fig.\ \ref{fig:model}. This setup, introduced in earlier studies of energy transport and multiple reflections \cite{Graciano2022}, leads to rich dynamical behaviour. As the particle travels between the walls, it can gain or lose energy depending on the phase of the moving base. Reflections at the infinitely high wall at $x = 0$ confine the particle, which undergoes repeated interactions with the oscillating region until its energy surpasses a critical threshold (set at $e = 1$), allowing escape. The resulting dynamics are conveniently described by a two-dimensional, area-preserving non-linear map in terms of dimensionless energy $e_n$ and phase $\phi_n$.

In dimensionless form, the discrete-time evolution of the system is described by the following non-linear map
\begin{align}
e_{n+1} &= e_n + \delta\left[\cos(\phi_n + i\Delta\phi_a) - \cos \phi_n\right],\\
\phi_{n+1} &= \left(\phi_n + i\Delta\phi_a + \Delta\phi_b\right) \mod 2\pi,
\end{align}where \( i \) is the smallest positive integer such that \( e_{n+1} > 1 \). The auxiliary phase increments are defined as
\begin{align}
\Delta\phi_a &= \frac{2\pi N_c}{\sqrt{e_n - \delta\cos(\phi_n)}},\\
\Delta\phi_b &= \frac{2\pi N_c r}{\sqrt{e_{n+1} - 1}}.
\end{align}

The effective control parameters of the systems are: $r$, the ratio between the fixed and oscillating regions; $\delta$, which determines the amplitude of the temporal modulation; and $N_c$, a dimensionless parameter controlling the characteristic time scale of the oscillations.

A Poincaré section is defined at the fixed position $x = (a + b)/2$, marking the end of each full cycle of motion across the well. This stroboscopic sampling captures the system's phase space in terms of $(e, \phi)$ and reveals its typical mixed-type character: regular islands embedded in a chaotic sea, separated by invariant curves. Crucially, the integer $i$ introduces a non-trivial dynamical feature: it quantifies the number of oscillation cycles required for the particle to accumulate sufficient energy to leave the oscillating region. Regions where $i > 1$ correspond to \textit{multiple reflections}, which are often associated with stickiness phenomena—transiently trapped trajectories that play a central role in the emergence of recurrent high-energy states.

\begin{figure}
    \centering
    \includegraphics[scale=0.35]{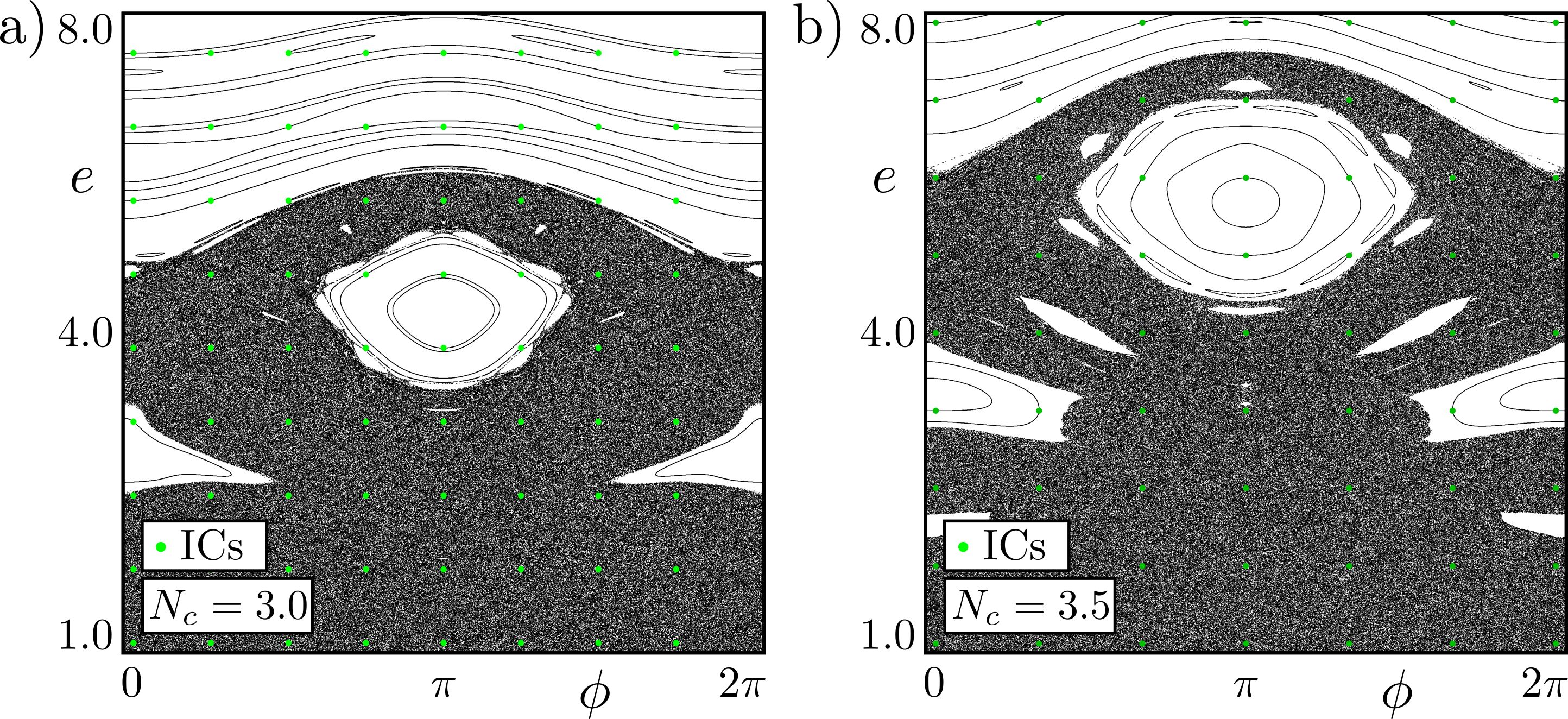}
    \caption{Characteristic mixed phase spaces of the time-dependent potential well model for $r = 1.0$ and $\delta = 0.5$. Panels show: (a) $N_c = 3.0$ and (b) $N_c = 3.5$. Green dots represent the set of ICs, uniformly sampled along $\phi \in [0, 2\pi]$ and $e \in [1.0, 8.0]$.}
    \label{fig:phase_spaces}
\end{figure}

Figure\ \ref{fig:phase_spaces} displays the phase space of the system in the plane $(\phi, e)$ for two representative values of the control parameter $N_c$. In both panels, the same ensemble of initial conditions is uniformly distributed along $\phi \in [0, 2\pi]$ and $e \in [1.0, 8.0]$. As a consequence of the Poincaré section defined at $x = (a + b)/2$, a mixed-type phase space emerges, composed by a chaotic sea surrounding periodic islands, and limited by invariant spanning curves.

By increasing the value of $N_c$, we observe a marked expansion of the chaotic regions in phase space. Specifically, the comparison between panels (a) and (b) illustrates how higher values of $N_c$ lead to the destruction of invariant tori and a more dominant chaotic sea. This behaviour is consistent with a decrease in the effective modulation frequency of the system, enhancing resonance overlap and promoting diffusion across phase space.

We explore the dynamical behaviour of many different initial conditions considering these both two values of $N_c = 3.0$ and $N_c = 3.5$. Although $r$ and $\delta$ also influence the underlying dynamics, we keep them fixed ($r = 1.0$ and $\delta = 0.5$) throughout the present analysis and defer a systematic exploration of all control parameters to future studies.

\section{Methodology}
\label{sec:methodology}

We employ Ensemble Recurrence Analysis (ERA) to systematically study the system's ICs. A trajectory is said to recur if, at some time $t_j$, it returns to the dynamical neighbourhood of a previous state at $t_i$ $(t_i < t_j)$, i.e.\ if $\boldsymbol{x}(t_j) \approx \boldsymbol{x}(t_i)$. Given a threshold distance $\varepsilon$, one can define the binary Recurrence Matrix (RM), whose elements $R_{i,j}$ are given by
\begin{equation}
    R_{i,j}(\varepsilon)=\left\{\begin{array}{ll}
    1, & \text{if} ~\Vert \boldsymbol{x}(t_i)-\boldsymbol{x}(t_j) \Vert < \varepsilon,\\
    0, & \text{otherwise}
    \end{array}
    \right.,
    \label{eq:RM}
\end{equation}where $\Vert\cdot\Vert$ denotes a suitable norm. Each entry equal to $1$ represents a recurrence, meaning that the system's state at $t_j$ lies within an $\varepsilon$-neighbourhood of the state at $t_i$. Since we focus on ensemble and finite-time recurrence analysis, we consider $0 < t_i < t_j \leq N$, where $N$ is the maximum number of iterations of the trajectory. Accordingly, the RM has dimensions $(N \times N)$.

A visual representation of the RM is known as the Recurrence Plot (RP). Typically, an RP displays the entries $R_{i,j} = 1$ as coloured pixels and the $0$ entries as white. RPs can exhibit markedly different structures depending on the orbit's underlying dynamics, which are entirely determined by the choice of its ICs in deterministic systems. Moreover, a single RP can reflect multiple dynamical regimes over time for a given trajectory. For instance, a chaotic trajectory initially trapped near a regular region may exhibit recurrence patterns resembling quasi-periodic motion. As the trajectory evolves, it may escape into a more irregular part of the phase space, producing a recurrence pattern characteristic of chaotic dynamics. In such cases, the RP will naturally display distinct temporal windows, each revealing different recurrence structures, until the final time $t = N$ is reached.

\begin{figure}
    \centering
    \includegraphics[scale=0.9]{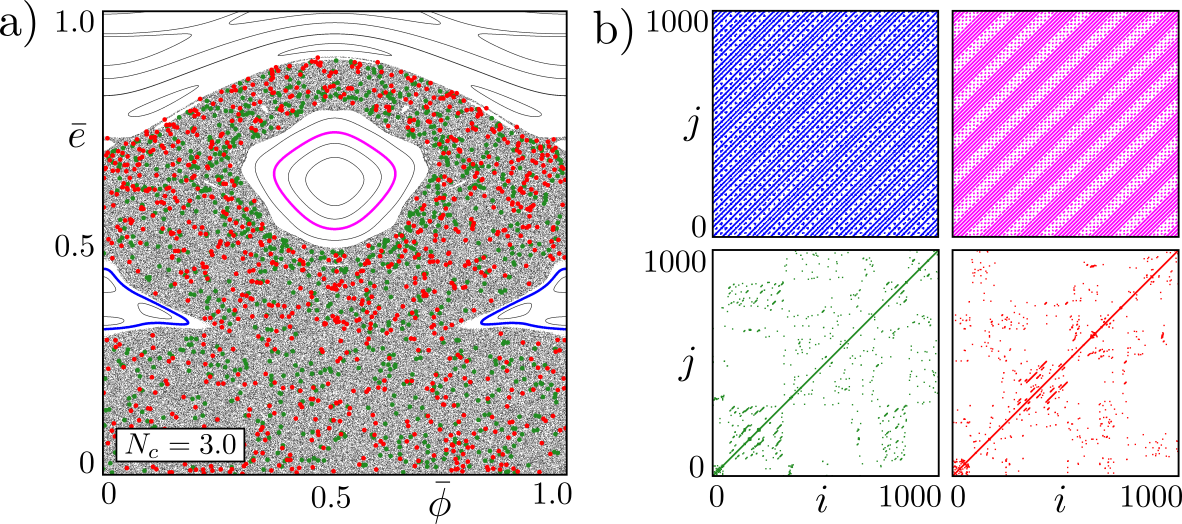}
    \caption{(a) Phase space of the time-dependent potential well for $N_c = 3.0$ showing four representative orbits in different colours. The projection is taken in the normalised coordinates $(\bar{\phi}, \bar{e})$, ensuring balanced contribution from both axes when computing distances for recurrence analysis. (b) Recurrence plots corresponding to the orbits in panel (a), plotted with the same colour code. The diagonal recurrence structure in the top panels indicates quasi-periodic motion, while the bottom panels reflect chaotic dynamics.}
    \label{fig:orbits_and_rps}
\end{figure}

To better illustrate the connection between the system's phase space and its recurrence behaviour, Fig.\ \ref{fig:orbits_and_rps} presents four selected trajectories and their corresponding RPs, chosen to highlight qualitatively distinct dynamical behaviours. Panel (a) shows the normalised phase space $(\bar{\phi}, \bar{e})$, where each coloured dot represents a sampled point from an individual orbit. The normalisation is essential to ensure that the Euclidean distances computed in Eq. \eqref{eq:RM} are not biased toward either the energy or phase coordinate, thereby providing a balanced representation in the RP. Panel (b) displays the RPs of these trajectories, using the same colour code. The two upper panels exhibit highly structured diagonal lines, characteristic of regular and quasi-periodic motion. In contrast, the bottom plots reveal chaotic features: the green RP shows intermittent recurrence patterns, while the red RP portrays dispersed and less coherent patterns typical of fully chaotic dynamics.

Figure \ref{fig:orbits_and_rps} is an example of the ability of recurrence analysis to differentiate between dynamical regimes and to capture subtle transient phenomena, such as stickiness, that may be hidden in traditional phase space visualisations.

Once the RP of a trajectory of interest is computed, many different measures can quantify and differentiate several aspects between different RPs. These are called Recurrence Quantification Analysis (RQA) \cite{MarwanWebber2015}. The simplest one is the Recurrence Rate ($RR$), which provides the percentage of recurrence points as follows
\begin{equation}
    RR = \frac{1}{N^2} \sum_{i,j=1}^{N} R_{ij}~,
    \label{eq:RR}
\end{equation}note that $RR = RR(N)$, i.e.\ the recurrence rate depends on the maximum iteration time considered for the evolution of the given trajectory.

\begin{figure}
    \centering
    \includegraphics[scale=0.27]{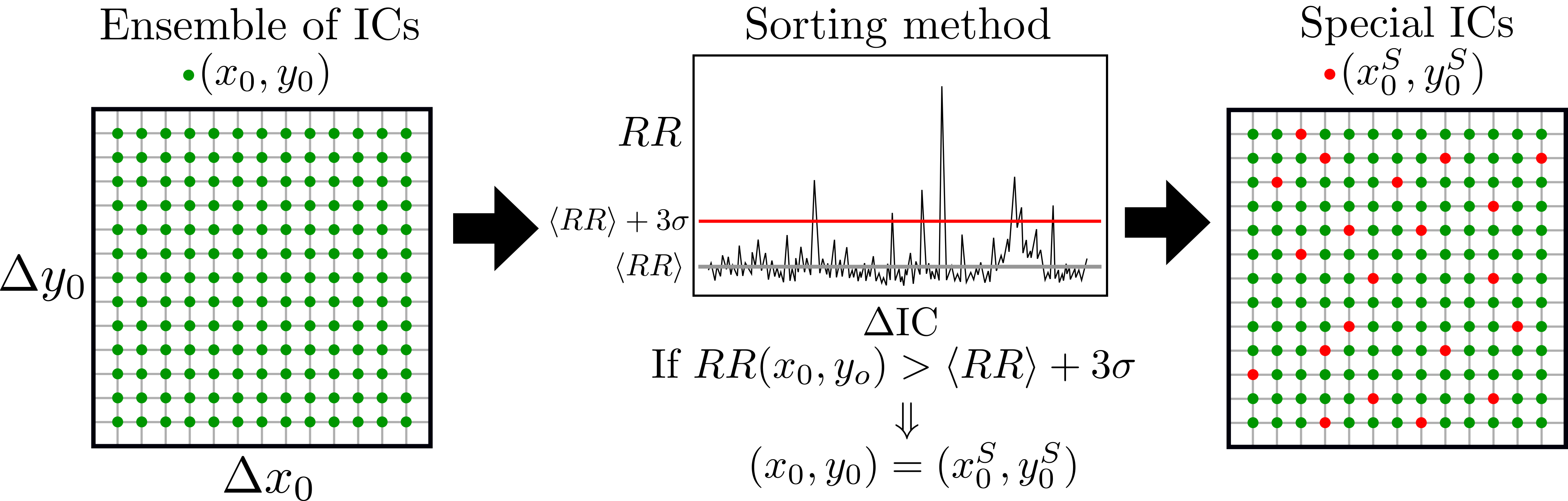}
    \caption{Schematic general example of the sorting procedure to detect the special ICs $(x_0^S, y_0^S)$ with the ERA methodology. The limit, depicted by the red line at the centre panel, can be defined by a statistical threshold e.\ g.\ 3 times the standard deviation $\sigma$ of the average recurrence rate $\langle RR \rangle$.}
    \label{fig:sketch_detect}
\end{figure}

To explore the system's dynamical landscape systematically, we adopt the Ensemble Recurrence Analysis (ERA) framework, initially introduced in \cite{Palmero2022}. As sketched in Fig.\ \ref{fig:sketch_detect}, we initialise a uniform grid of initial conditions across the normalised phase space and evolve each trajectory up to a fixed iteration time $N$. For each of these trajectories, the corresponding RP is constructed, and its $RR$ value is computed via Eq.\ \eqref{eq:RR}.

The collection of $RR$ values over the ensemble of ICs is then statistically analysed. In particular, we focus on identifying orbits that exhibit unusually high recurrence rates, these are typically associated with transient trapping near regular structures, i.e.\ the stickiness phenomenon. To detect such orbits, we apply a simple peak-detection method over the distribution of RR values across the ensemble. ICs corresponding to the upper limit of this distribution (i.e.\ statistically significant peaks) are selected as special ICs $(x_0^S, y_0^S)$.

This methodology allows us to isolate and study orbits that remain recurrent over extended time intervals despite exhibiting chaotic behaviour. As will be shown, such trajectories are not only dynamically rich but also statistically rare. Notably, they exhibit excursions to high energy states, with energies reaching approximately $e \approx 4$ and $e \approx 8$. These values are significantly high both in relation to the initial energy ($e_0=1.10$) and to the dimensionless amplitude of the potential well’s oscillations ($ \delta = 0.5 $), thereby characterising them as rare high-energy episodes in the system's evolution.

\begin{figure}
    \centering
    \includegraphics[scale=0.3]{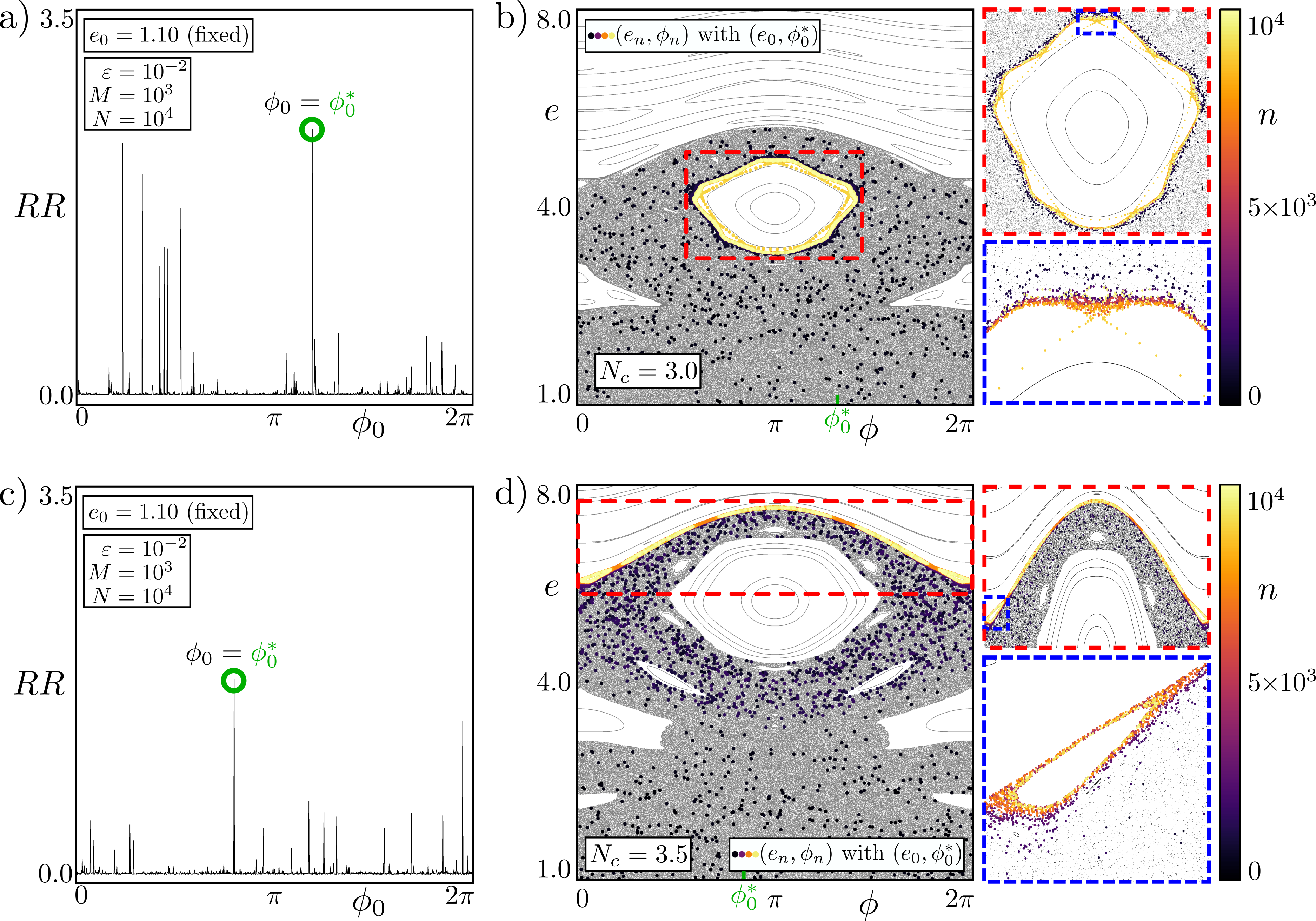}
    \caption{(a) Distribution of $RR(\phi_0)$ for $N_c = 3.0$; (b) Phase space for $N_c = 3.0$ showing the trajectory initiated at $(e_0,\phi_0^*)$, colour scale encodes the iteration time $n$; (c) Same as (a) for $N_c = 3.5$; (d) Same as (b) for $N_c = 3.5$. Insets in (b) and (d) show magnified views of regions where the trajectory exhibits transient trapping.}
    \label{fig:rr_phase_sticky}
\end{figure}

\section{Results}
\label{sec:results}

Our numerical results, obtained through the methodology presented in the previous section, reveal specific ICs that lead to special chaotic trajectories. These trajectories are characterised by high $RR$ and exhibit sustained quasi-periodic behaviour, frequently reaching elevated energy levels for extended periods. Despite originating from low-energy $e_0 = 1.10$, they confirm the existence of transient yet dynamically robust high-energy states.

Figure\ \ref{fig:rr_phase_sticky} shows the identification and dynamical consequences of high-RR trajectories. Panels (a) and (c) display the distribution of $RR(\phi_0)$ for an ensemble of trajectories with fixed energy $e_0 = 1.10$ and uniformly sampled $\phi_0 \in [0, 2\pi]$, evolved up to $N = 10^4$ iterations. A simple peak detection identifies the initial phase $\phi_0^*$ corresponding to the highest $RR$ value, marked by a green circle.

\begin{figure}
    \centering
    \includegraphics[scale=0.29]{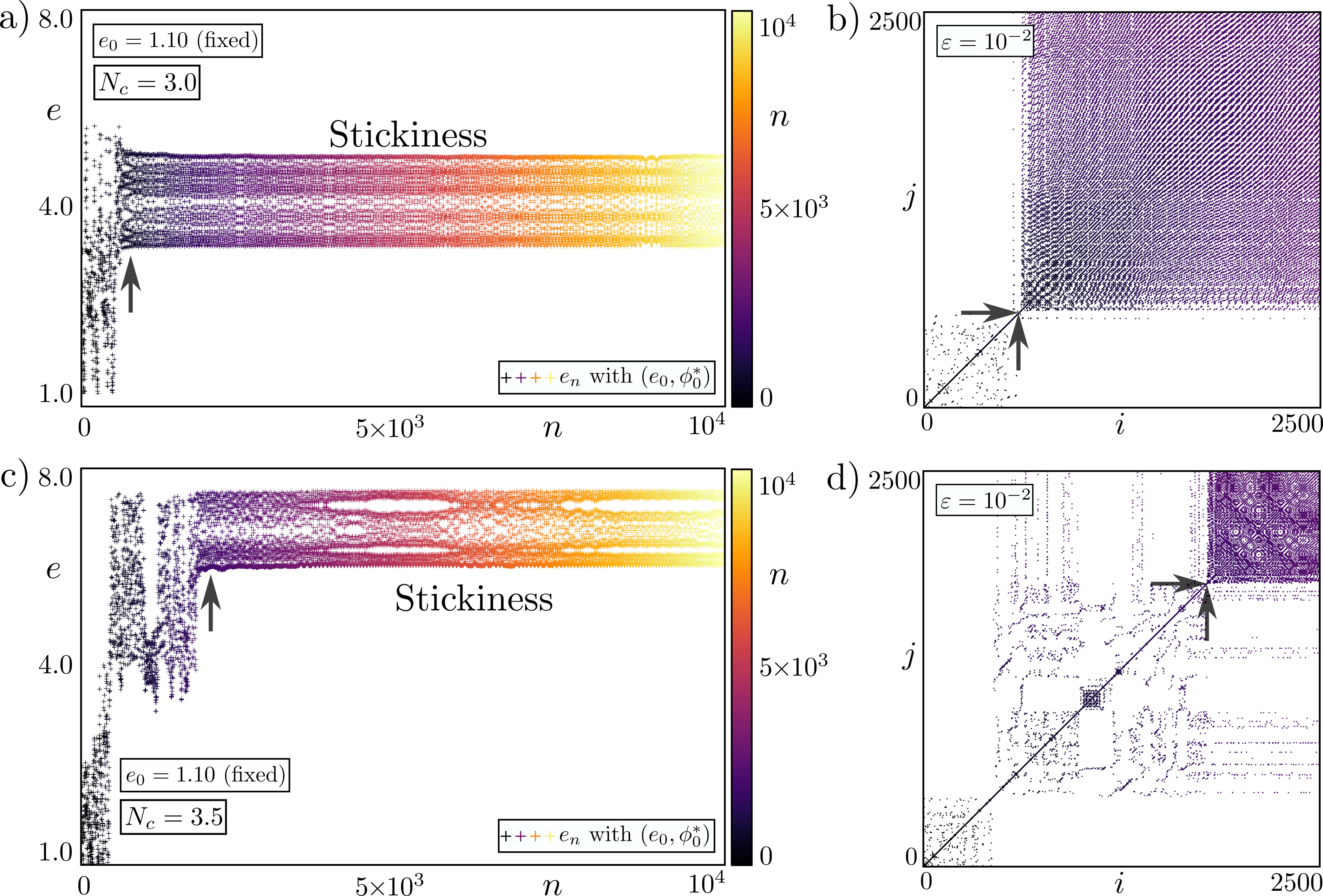}
    \caption{(a,c) Time series of the energy $e_n$ for the special trajectory initiated at $(e_0, \phi_0^*)$ with $e_0 = 1.10$, shown for $N_c = 3.0$ and $N_c = 3.5$; (b,d) Corresponding RPs constructed with threshold $\varepsilon = 10^{-2}$. The colour encodes iteration time $n$, following the same scale as in Fig.\ \ref{fig:rr_phase_sticky}, and the arrows mark the onset of stickiness.}
    \label{fig:time_sticky}
\end{figure}

The corresponding trajectories are shown in panels (b) and (d), projected onto the phase space $(\phi, e)$ for $N_c = 3.0$ and $N_c = 3.5$, respectively. The colour scale indicates the iteration time $n$, thereby revealing the temporal dynamics of energy diffusion. In both cases, the selected orbits begin by diffusing through the chaotic sea and later experience prolonged trapping near the boundaries of regular islands, marked in yellow and orange, as they enter the regions of stickiness for thousands of iterations. Insets in panels (b) and (d) present magnified views of these regions, highlighting how the trajectory oscillates along the border of invariant structures before eventually escaping (they do not escape until $10^4$ iterations, the maximum that we analysed). This behaviour is characteristic of stickiness in mixed phase spaces and is strongly correlated with high $RR$.

Figure~\ref{fig:time_sticky} presents the temporal dynamics and recurrence structure of the special trajectories identified through the ERA methodology. Panels (a) and (c) show the evolution of the energy $e_n$ for ICs $(e_0, \phi_0^*)$ with $e_0 = 1.10$, for the same two representative values of the control parameter, $N_c = 3.0$ and $N_c = 3.5$, respectively. In both cases, the trajectories exhibit a clear transition from initial chaotic diffusion to long-lasting trapping near regular structures, sustaining energy levels up to four and even eight times the initial energy. This behaviour provides a direct dynamical signature of stickiness. The trajectories become temporarily confined near the boundaries of stability islands, where energy fluctuations are suppressed and the motion mimics regular dynamics. This trapping phase is maintained over thousands of iterations, as reflected by the colour gradients and flat energy plateaus in the time series. Panels (b) and (d) display the corresponding RPs computed over the first 2,500 iterations. The structured diagonal patterns appearing after the initial segment (highlighted by arrows) confirm the onset of stickiness and the transition from fully irregular motion to a more correlated, quasi-regular regime. These structured recurrence domains coincide with the sustained high-energy states observed in the time series.

Together, these results demonstrate that the trajectories selected via the highest $RR$, although statistically rare, are dynamically significant, exhibiting prolonged trapping and sustained energy accumulation that would be otherwise difficult to detect through conventional phase space analysis.

\section{Conclusions and Perspectives}
\label{sec:conclusions}

We have shown that recurrence analysis, particularly through the ensemble framework, serves as an effective tool for identifying rare, highly recurrent chaotic trajectories that exhibit stickiness. Using a model based on a time-dependent potential well, we demonstrated that specific ICs, despite originating from low energy, may give rise to trajectories that remain trapped near regular structures for long time intervals, sustaining transient states of significantly high energy.

These results highlight a strong correlation between high recurrence rates and the dynamical manifestation of stickiness in mixed phase spaces. The approach successfully isolates dynamically rich orbits that would be difficult to detect using standard phase space visualisations. More broadly, this type of recurrence-based analysis opens a promising avenue to further investigate the stickiness phenomenon itself—potentially addressing open questions regarding its duration, finite-time dynamical stability, and the conditions under which such events may be sustained or enhanced. Further investigations could also examine the structural stability of these orbits under noise perturbations, as well as their statistical weight in long-term ensemble dynamics. The combination of recurrence-based diagnostics with classical tools may thus offer a broader and more quantitative understanding of rare dynamical features in complex systems.

In addition, our findings open the path for a more systematic exploration of the time-dependent potential well model. The idea would be further explore its control parameters, in order to identify regions of parameter space that favour the emergence of strong stickiness. Such an analysis could reveal configurations that are statistically more likely to support sustained high-energy states. This may prove particularly relevant for applications where rare or extreme dynamical behaviour plays a critical role, such as in transport phenomena, escape dynamics, or targeted control in Hamiltonian systems.

\section*{Acknowledgements}
M. S. Palmero acknowledges the financial support of S\~ao Paulo Research Foundation (FAPESP), grants n\textordmasculine\ 2023/07704-5 and 2024/22136-6. F. H. Graciano acknowledges the Federal Institute of Education, Science and Technology of Minas Gerais. E. D. Leonel acknowledges support from Brazilian agencies CNPq n\textordmasculine\ 301318/2019-0 and 304398/2023-3 and FAPESP grants n\textordmasculine\ 2019/14038-6 and 2021/09519-5. J. A de Oliveira acknowledges CNPq n\textordmasculine\ 304264/2025-3, 309649/2021-8 and 303242/2018-3.

%
%
\bibliographystyle{ieeetr}
\bibliography{main.bib}

\end{document}